\documentstyle[aaspp4,tighten]{article}
\begin{document}

\title{3C 295, A CLUSTER AND ITS COOLING FLOW AT Z=0.46}

\author{ D.M. Neumann}
\affil{ 
CEA/Saclay,
Service d'Astrophysique, Orme des 
Merisiers, B\^at. 709,
91191 
Gif-sur-Yvette Cedex, France
\\
Max-Planck-Institut f\"ur extraterrestrische Physik, Giessenbachstr. 1
D-85740 Garching, Germany,\\
email:ddon@cea.fr
}

\begin{abstract}
We present ROSAT HRI data of the distant and X-ray luminous
($L_x(bol)=2.6^{+0.4}_{-0.2}\times10^{45}$~erg/sec) cluster of galaxies
3C 295.
We fit both a one-dimensional and a two-dimensional isothermal $\beta$-model
to
the data, the latter one taking into account the effects of the point
spread 
function (PSF).
For the error analysis of the parameters of the two-dimensional model we 
introduce a Monte-Carlo technique.\\ 
Applying a substructure analysis, by subtracting a cluster model from
the data,
 we find no evidence for a merger, but we see a decrement in emission 
South-East of the center of 
the cluster, which might be due to absorption.\\
We confirm previous results by Henry \& Henriksen (1986) that 3C 295
hosts a 
cooling flow.
The equations for the simple and idealized cooling flow analysis presented here
are solely based on the isothermal $\beta$-model, which fits the data
very well, including the center of the cluster.
We determine a cooling flow radius of 60--120~kpc and mass accretion
rate of
$\dot{M}=400-900$~M$_\odot$/y, depending on the applied model and temperature 
profile. \\
We also investigate the effects of the ROSAT PSF on our estimate of
$\dot{M}$, which tends to lead to a small overestimate of this quantity if not 
taken into account. This 
increase of $\dot{M}$ (10-25\%) can be explained by a
shallower gravitational
potential inferred by the broader overall profile caused by the PSF, which 
diminishes the efficiency of mass accretion.
\\
We also determine the total mass of the cluster using the hydrostatic
approach.
At a radius of 2.1~Mpc, we estimate the total mass of the cluster 
(M$_{tot}$) to be $9.2\pm2.7\times10^{14}$M$_\odot$. For the gas to total mass 
ratio we get
M$_{gas}$/M$_{tot}$=0.17--0.31, in very good agreement with the results
for 
other clusters of galaxies, giving strong evidence for a low density
universe.
\end{abstract}


\section{Introduction}
In recent years the investigation of distant clusters of galaxies 
became a
key issue for the study of cosmological parameters as well as for 
structure formation. In addition, the comparison of the physical properties of distant
with nearby galaxy clusters can give useful insight into their
evolution.
\\
The distant cluster 3C 295, also known as Cl 1409+526 has been extensively
studied 
in optical wavelength bands: Dressler \& Gunn (1992) determined its
velocity 
dispersion to be
$\sigma=1300\,$km/sec in a study of seven distant clusters of galaxies. 
Thimm et al. (1994) in the search for line emission in galaxies
observed the cluster with an 
imaging Fabry-Perot interferometer and found the percentage of emission
line
galaxies  
in 3C 295 to be $40\pm11$\%. Smail et al. (1997) presented the
results
of a weak lensing analysis of 12 distant clusters, among them 3C
295.
The analysis they  present is based on data of the 
{\it Hubble Space Telescope}.\\
Henry \& Henriksen (1986) presented the analysis of {\it Einstein} data
of
the cluster. The authors conclude that there is strong evidence for 3C
295 
to host a cooling flow in the center.\\
Cooling flows generally occur in clusters in which the 
cooling time of the hot gas, the intracluster medium (hereafter ICM),
is less than the age of the cluster, so that the energy loss due 
to radiation cannot be neglected and mechanisms for the compensation
must 
occur. This was first found by Cowie \& Binney (1977), 
Fabian \& Nulsen (1977), and  Mathews \& Bregman (1978), based on the
findings 
of observations by Uhuru (Lea et al. 1973) -- for reviews see 
Fabian et al. (1984, 1991) and Sarazin (1986, 1988).
\\
In recent years a handful of other high redshift clusters also show indications
of hosting cooling flows in their centers. Most recently Schindler et al.
(1997)
reported on RXJ1347-1145 (z=0.45) having an extremely  massive cooling
flow with a mass accretion rate exceeding 3000~M$_\odot$/y. Fabian \& 
Crawford (1995) reported
on the detection of a cooling flow in IRAS P09104+4109 (z=0.442) with a
mass accretion
rates of about 1000~M$_\odot$/y, and Edge et al. (1994) found in Zwicky
3146
at z=0.29 a cooling flow with similar strength. 
\\
In this paper we aim to give more insight into the cooling flow and its 
dependence on the temperature structure of the ICM, the gas and dark
matter 
distribution and the overall dynamical state of 3C 295.
The outline is therefore as follows:
Sec.2 gives a brief description of the observation and data handling. In
Sec.3 we describe our spatial
analysis based on the isothermal $\beta$-model, which is followed by a
section describing our way of calculating the X-ray luminosity of the
cluster.
The cooling flow analysis is presented in Sec.5. 
Sec.6 describes the details of our employed mass analysis.
Discussion and conclusion follow hereafter.
\\
Throughout the paper we assume $H_0=50$km/sec/Mpc, $q_0=0.5$, and
$\Lambda=0$.

\section{X-ray observations}
3C 295 was observed with the ROSAT HRI for 29.6~ksec. Fig.\ref{fig:plot}
shows the resulting image of the cluster. We only use channels 2 to 
9 out of the 15 channels of the instrument to optimize 
the signal to background ratio as the neglected channels mostly have low
or 
negligible sensitivity (for details see David et al. 1997). 
With about 830 source photons within a radius of 8.3~arcmin  we obtain a 
countrate $0.028\pm0.001$s$^{-1}$ for the cluster. 

\section{Spatial analysis}

\subsection{Overall morphology}

The cluster is elongated in North-South direction with an axial ratio of 
0.78 (see below and Tab.\ref{tab:beta}). The emission is centrally 
peaked, which is a strong 
indication
for the existence of a cooling flow (see below). There is some isophot
twist 
visible in the contours in Fig.\ref{fig:plot}.  
The lines are more compressed in the South-East and more stretched in
the
North-West. We will discuss this feature and its significance later on
in more 
detail.
\\
There are no other clearly extended sources visible in the vicinity of the
cluster as
all other emission peaks in  Fig.\ref{fig:plot}, which have
at least two contour lines, are consistent with point
sources -- we checked whether these sources were possibly extended by
determining their
individual surface brightness profiles and comparing them to the on-axis
point spread function of the ROSAT HRI. 

\subsection{Surface brightness and isothermal $\beta$-model}

The radial surface brightness profile of the cluster can be seen in 
Fig.\ref{fig:surf}. 
To explore physical quantities of 3C 295 we make use of the isothermal 
$\beta$-model (Cavaliere \& Fusco-Femiano 1976, 1981; Sarazin 1986).
This
model enables us to convert the brightness profile of a cluster
analytically into a density profile of the gas.
The 
one-dimensional surface brightness profile of this model has the
following 
form:

\begin{equation}
S(r) = S_0(1+\frac{r^2}{a^2})^{-3\beta+1/2}+B,
\end{equation}

which translates into a density profile:

\begin{equation}
n(r) = n_0(1+\frac{r^2}{a^2})^{-3\beta/2},
\label{equ:density}
\end{equation}

where $a$ (the so-called core radius), $S_0$ (the central surface
brightness), $B$ (the background) and $\beta$ are all fit parameters.  
\\
There exists also a two-dimensional form of the
isothermal $\beta$-model which takes into account a possible elongation of
the
cluster. This model has the representation :

\begin{equation}
S(x,y) = S_0 \left(1 + F_1 +
F_2\right)^{-3\beta+1/2}+B; 
\end{equation}

\begin{displaymath}
\mbox{with: }
F_1 = \frac{(\mbox{cos}(\alpha)(x-x_0)+
\mbox{sin}(\alpha)(y-y_0))^2}{a_1^2};
F_2 =
\frac{(-\mbox{sin}(\alpha)(x-x_0)+\mbox{cos}(\alpha)(y-y_0))^2}{a_2^2}.
\end{displaymath}

We apply in the following the  one-dimensional and the  two-dimensional 
isothermal $\beta$-model to the data. The first model
has four fit parameters ($S_0$, $a$, $\beta$, $B$), the second eight fit 
parameters ($S_0$, $a_1$, $a_2$ - two core radii, one for the major axis,
one 
for the minor axis, $\beta$, $B$, $x_0$, 
$y_0$ - the position of the center of the cluster in x and y, and the
position angle $\alpha$). In the one-dimensional fit we predefine the
center
of the cluster by sorting the photons into concentric annuli centered on
the
peak of the X-ray  cluster emission. In the two-dimensional case, where
we look
 at 
pixels instead of annuli, this predefintion is not necessary and the
center of 
the
cluster emission is simultanously fitted together with the other 
parameters.
\\ 
The 
two-dimensional fit we apply here takes into account the effects of the
PSF of 
the ROSAT HRI by convolving each tested model with the PSF before
fitting. 
For the fitting we exclude 
serendipitous sources, which are clearly not connected to the X-ray
emission
of the cluster. The positions of these excluded sources can be seen in 
Tab.\ref{tab:exclude}.

\subsubsection{The one-dimensional fit}

For the one-dimensional fit we bin the data in concentric 
annuli with a width of 10 arcsec.
The results for this fit give a
core radius of $a=7.2^{+3.4\prime\prime}_{-3.3} \equiv
50^{+23}_{-23}$~kpc and a slope 
parameter of $\beta = 0.56^{+0.10}_{-0.07}$
(2 $\sigma$-errors). The reduced $\chi^2$ of the fit is 1.03, which
gives a very high confidence in the fit results. The fitted profile can be seen
in Fig.\ref{fig:surf}, and details of the other fit parameters
are given in Tab.\ref{tab:beta}. Reducing the width of the annuli, in
order to
better resolve the central part does not significantly change the fit 
parameters and the connected uncertainties.
\\
The derived quantities are in good agreement
with the results obtained by Henry \& Henriksen (1986), who analyzed
Einstein data and derived a core radius of $a=9^{+10\prime\prime}_{-4}$ 
and a $\beta=0.58^{+0.25}_{-0.08}$ (1 $\sigma$-errors).

\subsubsection{The two-dimensional fit}
For the two-dimensional fit we use
an image with a pixel size of $3''\times3''$, taking into account all
pixels located within a radius of 2.5~arcmin from the pointing position
at RA=$14^h11^m21.6^s$, Dec.=$52^d12^m00.0^2$ (J2000).\\
Our fitting procedures for both the one-dimensional and the
two-dimensional
case are based on $\chi^2$-statistics. This is ideal for
the one-dimensional case, in which we look at concentric annuli centered
on
the emission peak of the cluster, as there are always enough
counts per bin to assure that the Poisson distribution in each bin
is similar to a Gaussian distribution, which is required for
$\chi^2$-fitting. 
The
decrease in cluster emission outwards in this case is compensated by the 
increase in area of the annuli. However, in the 
two-dimensional
case, in which we use image pixels all of the same size, this 
compensation does not occur. Here the
counts per pixel, especially in outer regions, are too low to show Gaussian 
behaviour, but instead, show a classical Poisson distribution. To 
overcome the
problem 
here, we apply a small Gaussian filter ($\sigma= 3''$)
to the image
before fitting. This Gaussian filter changes the Poisson statistics into a
Gaussian shape. For the error determination in each pixel we can fully
account
for the Gaussian filtering by adding the corresponding errors of all 
contributing pixels 
quadratically (the exact formalism of the error determination can be 
found in Neumann \& B\"ohringer (1997)).
In this way we change the errors in countrate per pixel without
reducing the number of degrees of freedom. We do not lower the number of 
degrees of freedom
here,
as we would in the case of increasing the pixel size of the used image,
since we 
keep all the available spatial information.\\
To account for the 
Gaussian filtering, which causes the inferred profile of the cluster to
become shallower, we 
also apply the same Gaussian filter to the point spread function image with
which 
we convolve the fitted cluster model. To prove that with this procedure
we 
indeed are able to determine the precise shape of the cluster we
constructed
a simulated cluster image with known properties, similar to
3C 295,
and with an equal exposure time as our analyzed ROSAT HRI pointing. We
apply 
Poisson noise on this 
image and convolve it with the on-axis PSF of the ROSAT HRI. 
Subsequently we apply the Gaussian filter and fit the two-dimensional 
$\beta$-model to the simulated cluster image, taking into account
the (Gaussian filtered) PSF. The results can be seen in Tab.\ref{tab:simul}.
The 
intrinsic cluster model properties and the obtained fit
parameters agree very well within the errors, giving clear evidence that
our
method indeed gives correct results. \\
As $\chi^2$-statistic does
not allow an error of 0, we define the error of a pixel with value 0 to
have
an error of 1 (before filtering). Because of this necessary
modification, we are, unfortunately,  not able
to give a confidence level or a reduced $\chi^2$ value
of the fitted elliptical cluster model, and therefore base our error 
determination on a Monte Carlo-technique (see below).
Our fit results are the following: for the core radii we get 
$a_1= 4.2^{+2.3\prime\prime}_{-2.3} \equiv 29^{+16}_{-16}$~kpc and 
$a_2 = 3.3^{+1.8\prime\prime}_{-1.8} \equiv 23^{+12}_{-12}$~kpc, for the
slope
parameter we get $\beta=0.52^{+0.07}_{-0.07}$ (2$\sigma$-results). The
results 
for
the other six fit parameters can be seen in Tab.\ref{tab:beta} and
the profile can be seen in Fig.\ref{fig:surf}. One can
see that while $\beta$ is relatively unaffected by the PSF of the ROSAT
HRI
the same is not true for the core radius. The core radius changes when
the
PSF is taken into account and drops from 7.2~arcsec to 4.2~arcsec or 
3.3~arcsec, respectively. However, one should note that the errors of
this
parameter (which are all 2 $\sigma$ errors) show a large overlap. 
\\
 The approach of fitting after the application of a Gaussian filter in the 
two-dimensional case
has been already
successfully applied to other clusters, such as Cl0016 (Neumann \&
B\"ohringer
1997) and A2218 (Neumann \& B\"ohringer 1998).

\subsubsection{Error estimates for the two-dimensional fit}

For the error determination of the 2d-fit parameters we apply a
Monte-Carlo 
method for which we 
construct 100 artificial images by adding random Poisson noise on the
real 
cluster image and fitting each individual Poisson image in the same way
as 
the original image. For the errors  we use
two times the standard deviation of each fit parameter from the 100 
different fit results.

\subsubsection{Comparison of the fits}
The comparison of the two fits to the original data, the one taking into
account the effects of the PSF the other one not, can be seen in 
Fig.\ref{fig:surf}. In fact, apart from the central region, which has a
more
peaked emission when taking into account the PSF, the differences 
in the surface brightness profile are rather small. Therefore we
conclude
that the PSF might affect the cooling flow analysis (see below), but not
the
overall cluster properties at large radii. In order to see whether the 
difference in the fit results are partly due to ellipticity, we perform 
a one-dimensional fit, which takes into account the effects of the PSF.
The
results are identical to our fit result of the two-dimensional fit,
showing
that the differences are only due to PSF effects. 

\subsubsection{Effects of blurring}

In the above analysis we neglected possible errors in the aspect resolution 
of the instrument. In order to see whether the PSF is degraded because of
effects concerning the wobbling of the telescope or the reaquisition of
guide stars we compared the theoretical PSF of the telescope and HRI with
point sources in the field-of-view. The comparison of theoretical and actual
PSF gives somewhat different results, depending on the binning of the actual 
observed point sources
- This effect is most likely due to Poisson noise as all point 
sources are with a count rate well below 0.003 cts/sec very faint. - The 
brightest source has about 60 source counts (see Tab.\ref{tab:exclude}). 
 However, allsurface brightness  profiles of serendipitous sources are quite 
consistent with the theoretical PSF (see above).
If the aspect resolution is slightly degraded, it 
is not possible to assess the error it introduces and it is also not possible 
to correct for it, due to the lack of bright sources in the pointing
(for corrections concerning blurring see for example Morse 1994 and
 Harris et al. 1998).\\
This blurring can in principle effect our $\beta$-model fits and the resulting 
analysis. However, we do not expect the effects to play an important role.

\subsection{Search for substructure}

In order to search for possible substructure in the cluster we construct
a cluster model image from the best fit parameters of our
two-dimensional fit, which includes the background (the model can be seen in
Fig.\ref{fig:sym}). 
Subsequently we convolve the model with the PSF
of the ROSAT HRI and subtract the model from the original data. The
image of 
the so obtained residuals are shown in Fig.\ref{fig:resi}.\\
In this image we find positive residuals only below a significance level
of
three $\sigma$ above background and cluster emission, so that there is
no
strong evidence for a perturbation in the cluster potential, like a
subgroup falling onto the cluster. There is a decrement
in the residual map in the South-East of the cluster, which could in
principle 
be an indication of some disturbance. 
However, it could also be just
an indication of a region with higher intrinsic galactic absorption.
Unfortunately, surveys of the 21-cm line do not (yet) provide the
necessary
spatial resolution to confirm or rule out a strong gradient in galactic 
absorption in the line-of-sight of the cluster.\\
The fact
that this is the only evidence that the cluster deviates from the
applied
cluster model weakens the case for a perturbation in the potential. 
To see whether this decrement might be due to an offset between the true 
cluster center and cluster model (which is very unlikely as the peak of 
emission in 
Fig.\ref{fig:plot} and the cluster  center in Fig.\ref{fig:sym} coincide
very 
well), we shift the center of the subtracted cluster 
model several arcseconds in the direction NW of the center. The decrement
remains
after the shift, indicating that this deficit in emission might indeed
be real.
\\
Finally, we conclude that, apart from this decrement, there is no further 
evidence for the cluster being dynamically young. It does not seem that the 
cluster is
suffering or has recently suffered from a major merger phase, which
is in agreement with the cluster hosting a cooling flow -- normally a strong
 indication for a relaxed cluster. This fact is strengthend by the high
confidence level of the one-dimensional fit, which one does not expect
in case
of a non-relaxed cluster.
However, we want to stress that our 3C 295 exposure has limited 
signal-to-noise, and that future X-ray telescopes, such as AXAF or
XMM will give further information due to higher sensitivity and lower 
background.

\subsection{Central electron density}

To calculate the central electron density of the cluster we employ the
results 
of the isothermal $\beta$-model. Using the one-dimensional model 
as input we calculate a central electron density of 
$n_{e0}=0.040^{+0.007}_{-0.005}\,$cm$^{-3}$.
For the two-dimensional $\beta$-model we use as mean core radius the 
geometrical mean of $a_1$ and $a_2$, giving $\bar{a}=3.7''$. The derived 
$n_{e0}$ in this case is  $n_{e0}=0.084^{+0.074}_{-0.049}$cm$^{-3}$.
The errors are hardly dependent on the 
temperature due to the ROSAT energy band (see for example Jones \&
Forman 1992;
B\"ohringer 1994)
but are dependent on the intrinsic
shape parameters of the isothermal $\beta$-model. The errors are larger
for 
the two-dimensional case, as we take into account the errors of the two
core radii, the $\beta$ and the $S_0$, while for the 1-d case we only
take
into account the error of the one core radius and $\beta$.

\section{X-ray luminosity}
The ROSAT/HRI source countrate of 3C 295 is  
$0.028\pm0.001\,$sec$^{-1}$. To determine this countrate we subtract from
the 
original data the background rate determined from the spherical
symmetric 
$\beta$-model fit and exclude serendipitous sources not belonging to the 
cluster.\\
Assuming that the cluster temperature lies between 
kT=$7.1^{+2.1}_{-1.3}$~keV, the result by Mushotzky \& Scharf (1997)
based on
ASCA data, we
obtain an X-ray rest frame luminosity of 
$L_x(0.1-2.4$~keV$)=1.01^{+0.05}_{-0.05}\times10^{45}$erg/sec in the
ROSAT 
band,
corresponding to a bolometric luminosity of
$L_x(bol)=2.6^{+0.4}_{-0.2}\times
10^{45}$erg/sec. 
The error on the luminosity is a combination of both countrate statistics
and the uncertainty in temperature. For the model we use thermal
Bremsstrahlung,
and a galactic hydrogen column density of
$n_H=1.34\times10^{20}$cm$^{-2}$
from 21cm-line measurements (Dickey \& Lockman 1990).
Our estimate for $L_x(bol)$ is in reasonably good agreement with the
result
of Mushotzky \& Scharf (1997), who obtain an X-ray luminosity of 
$L_x(bol)=1.9\times10^{45}$~erg/sec, when transformed to $q_0=0.5$. 
The discrepancy  in the luminosity estimate can have several sources. First of 
all the presence of colder gas in the cooling flow region might lower the
effective temperature in the ROSAT band, which is softer than the ASCA band.
Secondly, the discrepancy in the luminosity estimates might be due to
the large PSF of the ASCA instruments, 
leading normally to an underestimate of the countrate.\\
Our results are also in agreement with the results of Lea \& Henry
(1988),
who determined an X-ray luminosity of $L_x(0.5-4.5$keV$)=9^{+0.7}_{-0.7}
\times 10^{44}$erg/sec on the basis of Einstein data.

\section{Cooling flow analysis}

\subsection{Cooling time and cooling flow radius}
In central parts of clusters of galaxies the time in which the ICM cools from
several $10^7-10^8$K to temperatures several orders of magnitude less
 can be lower than the age of the cluster. In these regions
the energy loss due to radiation is therefore not negligible and mechanisms 
for the compensation must occur (for details see for example Fabian et al.
1991).
To give an estimate out to what radius the energy loss is not negligible
one
often makes use of the so-called cooling time $t_{cool}$, which is
defined
as the enthalpy of the ICM divided by the energy loss due to radiation
in
X-rays, {\it i.e.}
\begin{equation}
t_{cool} = \frac{\frac{5}{2}nk_bT}{n_en_h\Lambda(T)}.
\end{equation}
Here, $\Lambda$ is defined as cooling rate and is only dependent on
metallicity and
temperature. Typical values for $\Lambda$ are in the range of 
$1-3\times10^{-23}$erg s$^{-1}$cm$^6$.
Inserting our results from the isothermal $\beta$-model together with
some
estimates on $\Lambda$ from B\"ohringer \& Hensler (1989) (see also 
Tab.\ref{tab:lambda}), we can calculate 
the cooling time as function of radius. The cooling flow radius is then
defined
as the radius at which the cooling time is equal the cluster formation
time,
which depends on cosmological parameters. Assuming the cooling time to
be
$t_{cool}=7.4\times10^9$yr, which corresponds to the age of the cluster
at a 
redshift of z=0.46, we get cooling flow radii corresponding to
60--100~kpc, depending on the $\beta$-model and on the temperature we apply. 
For a 
comparison with nearby clusters, if we neglect the fact that we see the 
cluster 
in a younger stage than nearby ones, and assume a cooling time of 
$t_{cool}=10^{10}$~y, we
get cooling flow radii in the range of 70--120~kpc.

\subsection{Mass accretion rates}

If we assume that the only efficient mechanisms which compensate the
energy loss due to
radiation in the center of the cluster are mass accretion and cooling
of the ICM, we can write for the energy balance 

\begin{equation}
n_e(r)n_h(r)\Lambda(T) = \frac{\dot{M}(r)}{4\pi r^2}\left(\frac{5k}{2\mu
m_p}
\frac{dT}{dr}+
\frac{d\Phi(r)}{dr}\right),
\label{equ:balance}
\end{equation}
where $\Phi$ is the gravitational potential.
The left hand side of this equation is the energy loss due to radiation
as 
function of radius, the right hand side shows the mechanisms of
compensation: 
cooling and accretion. 
Now, we can insert our results of the isothermal 
$\beta$-model in equ(\ref{equ:balance}) and calculate the mass
accretion. To 
determine the gravitational potential we can use the hydrostatic
equation 
\begin{equation}
\frac{dP}{dr}=-\rho_{gas}\frac{d\Phi}{dr}.
\label{equ:hydro}
\end{equation}
Assuming that the ICM has the same properties as an ideal gas we can, in 
combination with the isothermal $\beta$-model, calculate the
gravitational
potential of the cluster and with it its total mass as function of
radius, {\it i.e.}
\begin{equation}
\Phi(r) = \frac{GM(r)}{r} = -\frac{rk_bT}{\mu m_p}\left(\frac{1}{n}
\frac{dn}{dr}+\frac{1}{T}\frac{dT}{dr}\right),
\label{equ:masse}
\end{equation}
where $M(r)$ is the total mass within radius $r$.
We employ the isothermal $\beta$-model for the gas density
distribution  even though the cluster might
not be isothermal as the emission in the ROSAT energy band (0.1-2.4~keV)
depends only weakly on the temperature of the ICM (see above).
We would like to stress here that we use the isothermal $\beta$-model only as 
fit function to deproject the surface brightness profile into a density 
profile of the ICM and that possible temperature gradients or the presence of
multiphase gas can be neglected
as the measured count rate in the ROSAT band hardly depends  on cluster 
temperature.
The use of any other analytical function instead of the $\beta$-model
for deprojecting would not increase
the reliability of the calculated density profile, as we do not have any
information on the temperature distribution. Deprojecting
numerically by subtracting shell by shell, as is done for nearby clusters
cannot be undertaken here, as our data have too poor statistics.
\\
We also want to note that our analysis presented here describes and
assumes idealized circumstances, for example 
that either the cooling flow region
is isothermal or has a constant temperature gradient.

\subsubsection{The isothermal case}
Assuming for now the stronlgy idealized case that the cluster is isothermal, 
and inserting 
equ(\ref{equ:density}),
equ(\ref{equ:hydro}) and equ(\ref{equ:masse})
in equ(\ref{equ:balance}), we can directly
calculate the mass accretion rate from
\begin{equation}
\dot{M}= \frac{4\pi \mu m_p a^2}{3k_bT\beta}n_{e0}n_{h0}\Lambda(T)r
\left(1+\frac{r^2}{a^2}\right)^{-3\beta+1}.
\end{equation}
With this equation one can see the dependence of $\dot{M}$ on $r$, $a$,
and
$\beta$. In case
that the cooling flow radius is smaller than the core radius $a$, or in
case
that the $\beta$ value is very small ($\beta \leq 0.33$), the 
mass accretion rate is proportional to $r$, since then the term 
$\left(1+\frac{r^2}{a^2}\right)^{-3\beta+1}$ almost stays constant. 
This is
in agreement with the findings of other authors, e.g. Thomas et al.
(1987), and
Fabian et al. (1991), Neumann \& B\"ohringer (1995). However, as in the
case of 3C295 the core 
radius is between 25~kpc and 50~kpc and therefore  smaller than the cooling 
flow radius, 
and as $\beta \ge 0.33$,
the term $\left(1+\frac{r^2}{a^2}\right)^{-3\beta+1}$
changes significantly and therefore influences the shape of the mass
accretion rate
as a function of radius. The results can be seen in Fig.\ref{fig:cf1} 
(for the isothermal $\beta$-model taking into account the PSF) and 
Fig.\ref{fig:cf2} (for the isothermal $\beta$-model neglecting the
effects
of the PSF). The maximal mass accretion rate is in the range 550--900
M$_\odot$/y, depending on the assumed temperature. In the isothermal
case the 
mass accretion rates rise with radius up to 50~kpc and then stay more
or 
less constant. 

\subsubsection{The case of constant temperature gradient}
It is generally found that the temperatures in cooling flow regions are
lower
than the overall cluster temperature. Often a temperature gradient 
in the cooling flow region can be 
observed. If we assume that there exists a constant temperature gradient
in the cooling flow region of 3C 295, we can rewrite
equ(\ref{equ:balance}) as
\begin{equation}
\dot{M} = \frac{4 \pi r^2 \mu m_p n_{e0} n_{h0}}{3k_b}
\frac{\left(1+\frac{r^2}{a^2}\right)^{-3\beta}\Lambda(T)}{
\frac{r(grad_T r+T_{center})\beta}{a^2\left(1+\frac{r^2}{a^2}\right)
}+\frac{grad_T}{2}},
\label{equ:grad}
\end{equation}
where $grad_T$ is the temperature gradient one infers and
$T_{center}$
is the central temperature.
In equ(\ref{equ:grad}) 
we take into account that a temperature gradient changes the shape of
the 
gravitational potential (see equ(\ref{equ:masse})).
 We assume that the gas is
still in hydrostatic equilibrium and that the hydrostatic equation 
equ(\ref{equ:masse}) is still applicable, assuming a single phase ICM.
One way 
to see this: while
the gas is cooling down its density distribution changes accordingly to
fulfill equ(\ref{equ:masse}).
For our calculation of mass accretion rates we
assume the temperature at the center to be 2~keV and at a radius of
120~kpc 
to be either 5.8, 7.1, or 9.2~keV -- the range of temperatures
representing the error in the determination
of the overall temperature of 3C 295 by Mushotzky \& Scharf (1997). For 
simplification
we assume that $\Lambda(T)$ is linear in temperature between 2 to
10~keV. 
The adopted values for $\Lambda$ can be seen in Tab.\ref{tab:lambda}.
For intermediate temperatures we interpolate linearly. The results of our
calculations
are shown in Fig.\ref{fig:cf1}, and Fig.\ref{fig:cf2}, respectively. 
Applying a temperature gradient lowers the mass accretion rates and
changes 
the shape of the mass accretion rate as a function of radius. The maximum 
mass accretion rate occurs close to the core radius of either model, being
in the range of 400--600~M$_\odot$/y. After its maximum the mass
accretion rate
drops at larger radii -- the mass accretion rate at 120~kpc varies
between 300--500~M$_\odot$/y.

\subsection{Comparison with previous work}
Already Henry \& Henriksen (1986) analyzing Einstein IPC and HRI data 
suggested that 3C 295 hosts a cooling flow in the
center. The authors calculated a mass accretion rate 
($\dot M$) of about 145 M$_\odot$/y. The determination of 
$\dot M$ was based on pointlike excess emission over the best
fit
isothermal $\beta$-model in the center. As one can see in
Fig.\ref{fig:surf},
we do not see excess emission in the center. The isothermal
$\beta$-model
coincides very well with the cluster emission in the center. This
suggests
two possibilities: the first one being that there was an active
point source in the center of the cluster, while it was exposed with the
Einstein HRI, and this source was not anymore active when exposed with
the 
ROSAT HRI. The second possibility is that because Henry \& Henriksen (1986)
simultanously fitted a point source and an isothermal $\beta$-model to
the data, this led to an overestimate of the central point source
resulting from an underestimate of the emission from the cluster itself. A 
weak indication
for this 
is the fact that Henry \& Henriksen obtain a core radius of 
$a=9^{+10\prime\prime}_{-4}$ and a $\beta=0.58^{+0.25}_{-0.08}$ -- both
values being 
higher than our best fit results. However, the overall agreement between 
Einstein and ROSAT data is very encouraging. The higher
fit values for core radius and $\beta$ obtained by Henry \& Henriksen
(1986)
cause the estimated cluster profile to be flatter in the center than in our 
fitted 
models.\\

\subsection{Contamination by radio or point sources emitting in X-rays}
The cluster of galaxies 3C 295 is known to host a luminous radio source
in
the center (Akujor et al. 1994 and references therein). The source
consists of 
two radio lobes
with a separation of about 5~arcsec. The extent of the lobes is each
about
2~arcsec. In principle the spatial resolution of the HRI is sufficient
to resolve these lobes, if they emit in X-rays. However the Poisson
statistics,
which is the factor which really limitats the spatial resolution here, 
makes a detection of their
separation difficult. It is also not clear 
whether these lobes
lead to an enhancement in X-rays or diminish the X-ray brightness in
this area.
An example of radio lobes dimming the X-ray luminosity in the
center of a cluster was shown by B\"ohringer et al. (1993) for the
Perseus 
cluster or Schindler \& Prieto (1997) in the case of A2634.\\
Strong 
evidence that these sources are not very prominent and do not
obscure the X-ray surface brightness profile is the fact that we yield a
very
high confidence level for the isothermal $\beta$-fit with a reduced 
$\chi^2$=1.03. The good agreement between fit and real data is difficult
to 
explain in the presence of additional sources in the center, whether
they 
enhance or diminish the X-ray brightness, as they affect the overall
fit.\\
Also a single point source in the center can be ruled out by the same
argument. It seems very unlikely that a central point source would have
exactly the ``right'' emission to fit the model. Also, in this unlikely
scenario, a central source would affect the profile only
out to 
about 3--5~arcsec -- the size of the PSF. As our calculated cooling flow
radius 
is larger than this ($12''-17''$), a single central point source cannot explain
the X-ray peak in the center.\\
Another test to see whether a central point source can influence the 
surface brightness profile in the center of the cluster is to perform an 
isothermal $\beta$-model fit (one-dimensional) neglecting the innermost
bin
of the radial profile (see Fig.\ref{fig:surf}). This central bin 
should contain almost all emission from a central point source, if 
present. The result of this fit leads to an even more centrally peaked
cluster 
model, with a best fit result for the core radius of about 1.5~arcsec, a 
$\beta=$0.54 and a very high central surface brightness of 
$S_0=8.3\times 10^{-4}$sec$^{-1}$ arcsec$^{-2}$. 
However, this fitted model has much larger error bars than the 
original one-dimensional fit, and the uncertainties in core radius and 
$\beta$ of the original fit are very well encompassed in the error
estimates 
of the modified fit.
\\ 
Putting this all together, we
conclude that if there are other  individual X-ray sources present in the 
center of the cluster, then their effect on the
cooling flow analysis is neglible, and they are unlikely to lead to a
dramatic
overestimate of the mass accretion rates.

\section{Mass analysis}
To determine the total and the gas mass of the cluster we utilize the
results 
of the isothermal $\beta$-model fitting for the gas density profile. We
assume that the temperature throughout the whole cluster lies between 
5.8-9.2~keV, which are the
results obtained by Mushotzky \& Scharf (1997) analyzing ASCA data. 
For the mass determination we assume spherical symmetry and hydrostatic
equilibirum and under these assumptions we can use equ(\ref{equ:masse}). 
These assumptions are justified and lead to reliable estimates for the
mass
determination of clusters of galaxies as shown by Schindler (1996) and
Evrard et al. (1996), who analyzed hydrodynamic simulations under these
assumptions and compared the obtained results with the intrinsic
properties.
For the
determination of the total mass we perform a Monte-Carlo analysis
(Neumann \& B\"ohringer 1995), which
calculates possible random temperature variations within the above
given
limits. For the stepwidth of these temperature variations we use 300~kpc.
As the uncertainties in temperature by far exceed the uncertainties in
the
density profile we neglect here the errors of the isothermal
$\beta$-model,
which introduce errors in the density distribution of the gas.\\
As a smoothing parameter, to avoid strong and unphysical 
oscillations of the temperature
profiles, we use $\Delta T=0.3$~keV, which means that each temperature at
a 
certain radius has to lie within 0.3~keV of the adjacent step radius. the temperature of the next inner step
radius T$\pm0.3$~keV. The results can be seen in Fig.\ref{fig:masse} and
Tab.\ref{tab:masse}. 
Calculating the total mass of the cluster we get
M$_{tot}=9.2\pm2.7\times
10^{14}$M$_\odot$ at a radius of 2.1~Mpc. The result is slightly dependent
on
the employed $\beta$-model. For the gas to total mass ratio we get 
M$_{gas}$/M$_{tot}=0.18^{+0.08}_{-0.04}$ for $\beta=0.56$
and M$_{gas}$/M$_{tot}=0.22^{+0.09}_{-0.05}$ for $\beta=0.52$. A summary
of the results is given in Tab.\ref{tab:masse}.

\section{Discussion}

\subsection{Cooling flow and $\beta$-model fitting}

Despite the fact that 3C 295 hosts a cooling flow, we do not see
evidence for 
central excess emission above the fitted isothermal $\beta$-model as seen in 
almost all other cooling flow clusters. Normally, 
fitting an isothermal $\beta$-model to the X-ray data of these kind of
clusters
leads to results which underestimate the central emission. However,
most
of these studies are based on nearby clusters, which can be traced out
to 
larger radii than 3C 295. We can trace the X-ray emission of 3C 295 only out 
to a radius of 2.0-2.5 arcmin (see Fig.\ref{fig:surf}), which corresponds
to 
900-1100~kpc. Therefore we are unable to see the outer parts of the
cluster and our fit is only based on the central parts. Fits of this
kind, where one only takes into account the center of a cooling flow
cluster, generally lead to smaller core radii and $\beta$'s. Our values for 
core radius and
$\beta$ are also relatively small, when compared to other clusters,
which
generally have core radii ranging between several 100~kpc and $\beta$'s,
which
lie between 0.6 and 0.9 for rich and X-ray luminous clusters of
galaxies.
Therefore we conclude that the reason we do not see a central excess
lies solely in the fact that we see and fit only the central part of 3C 295.
For an anology, we more or less see only the tip of the iceberg.
\\
Future 
X-ray
instruments such as aboard on AXAF or XMM, which have higher sensitivity
and 
lower 
background than the ROSAT/HRI will enable us to trace distant clusters
out
to much larger radii.

\subsection{The effects of the PSF on the cooling flow analysis}

We determine two different isothermal $\beta$-models in our spatial
analysis: one takes into
account the effects of the PSF and ellipticity, and gives a $\beta=0.52$
and
a small core radius ($\bar a=3.7$~arcsec), while in the other 
one neglects PSF effects which gives $\beta=0.56$ with a larger
value for 
the core radius ($a=7.2$~arcsec).\\
Comparing the results of the subsequent cooling flow analysis of the two 
models, one can see that in the $\beta=0.56$ case we overestimate the
cooling flow radius and the mass accretion rates (see Fig.\ref{fig:cf1}
and Fig.\ref{fig:cf2}). 
The reason why we determine a larger cooling flow radius when using the 
$\beta=0.56$ model instead of the $\beta=0.52$ model lies in the fact that
the inferred gas density profile shows a less steep drop in the first case.
This comes from 
the fact that the PSF makes the appearance
of the profiles more shallow.\\
Even though the calculated central electron density is lower in
the $\beta=0.56$ case, the densities of the gas
are equal in both models at a radius of 60~kpc, and at larger radii the 
$\beta=0.56$ model gives higher 
values for the gas density than the $\beta=0.52$ case.
The consequence is that the cooling times in the $\beta=0.56$ case at
radii larger than 60~kpc are smaller than the ones of the other model,
due to the fact that $t_{cool}\propto n^{-2}$.
As the cooling flow region is determined by its cooling time being less
than a
certain time interval, the cooling flow radius is, of course, bigger in
the 
$\beta=0.56$ case. The overestimate of the cooling flow radius is of the
order
of 20\%, but does not have a big influence on the resulting maximal mass 
accretion rate, as at outer regions the mass accretion rate stays
almost 
constant.\\
The overestimate of the mass accretion rates in the $\beta=0.56$
case again
results from a shallower profile, in this case from the gravitational
potential.
A shallower potential stands for a smaller potential gradient  and
this
means that more gas must flow into the center to compensate the energy
loss
due to radiation (see equ(\ref{equ:balance})). The overestimate of
the mass accretion rate due to PSF effects is of the order of 10--25\%,
which 
is not very high 
given all the other uncertainties of the physical properties, which
enter
the determination of the mass accretion rates.

\subsection{Mass accretion rates and temperature gradient}
Introducing a temperature gradient in the center of the cluster reduces
the
resulting mass accretion rates (see Fig.\ref{fig:cf1} and
Fig.\ref{fig:cf2}).
This shows how sensitive the determination of mass accretion rates is on
the temperature structure of the ICM. With a temperature gradient the
gas has another
mechanism 
to compensate for the energy loss due to radiation (we assume here the case 
of a homogenous cooling flow): while the gas is
falling 
into the center it not only looses potential energy but also kinetic
energy 
due to cooling.
However, assuming that the hydrostatic equation is valid in this region,
and assuming that multiphases in the gas are negligible, a temperature
gradient
also causes the gravitational potential to be more shallow 
(see equ.(\ref{equ:masse})).
Due to this shallower potential the energy supported by loss of
potential
energy of the gas falling inwards is not as efficient as in the
isothermal 
case.
However, as the overall mass accretion rates drop, the effect of cooling 
compensates easily the effect of a shallower gravitational potential.\\
We only want to stress here that our assumed models for the cooling flow
are simple, as we, for example, neglect the presence of a multiphase gas in the
cooling 
flow region. However, as 3C 295 is a very distant cluster, a fact which
makes 
a more detailed analysis of the cooling flow almost
impossible at the moment,
we only want to give  an idea of what mechanisms occur in the cooling
flow 
region of this cluster, and how sensitive the analysis is to the
observational
uncertainties.

\subsection{Correlation mass accretion rate and cooling flow radius}
Comparing the determined mass accretion rate and the calculated cooling
flow
radius with other clusters, for example the sample of White et al.
(1997),
one can see that 3C 295 is relatively outstanding. It shows a very high
mass 
accretion rate for its size of cooling flow radius. Following the
fit 
results of White et al. (1997), the cluster should have a mass
acccretion rate 
of about 60~M$_\odot$/y with a cooling flow radius of roughly 100~kpc --
alternatively, with
its mass accretion rate it should have a cooling flow radius of about 200~kpc.
However,
the difference between the calculation of White et al. and here is that
we 
neglect inhomogenities in the cooling flow, in which material can cool
out
completely and provides additional energy resources for radiation.
But it is
very unlikely that this effect causes the mass accretion rate to drop by
a 
factor of ten. The discrepancy lies more likely
 in the fact that White et al. look
at clusters at a redshift z$<0.2$, while 3C 295 lies at z=0.46. It might
be 
that the discrepancy is some sort of evolutionary effect. This idea is 
strengthened by another high redshift cooling flow cluster 
RXJ1347-1145 (Schindler et al. 1997) at z=0.45 with a mass accretion
rate of 
$\dot{M}>3000$M$_\odot$/y, and a cooling flow radius of 200~kpc, which
also does 
not fit in the relation of White et al.. Also the results of two other 
distant clusters, Zwicky 3146 (z=0.291) and Abell A1835 (z=0.252) by 
Edge et al. (1994) and Allen et al. (1996a)
show mass accretion rates also somewhat too large for their cooling flow
radii
to match the result by White et al..
It is therefore important to look at other
high redshift cooling flow clusters, to see 
whether this is indeed an effect depending on redshift, or whether it is
just that these clusters are exceptional in their properties.
 
\subsection{The baryon fraction in 3C 295}

The determined gas to total mass ratio in 3C 295 is 
M$_{gas}$/M$_{tot}\sim$0.22 -- similar to the findings in other clusters.
As the 
gas to total mass ratio
is a limit on the baryon fraction in clusters, which is thought to be
representative for the universe in total (based on numerical
simulations,
see for example Evrard 1990; Cen \& Ostriker 1993), 
this approach can be used to measure the fraction of baryon to dark
matter 
density in the entire universe (the mass in galaxies in clusters is
only a 
small fraction of the mass of the ICM).
 As studies of 
primordial
nucleosynthesis give precise values of $\Omega_b$ (the ratio of baryon
to
critical density of the universe), with $\Omega_b
h_{50}^{2}=0.05\pm0.01$
(see for example Walker et al. 1991) this can be directly used to
determine 
$\Omega_{tot}$, the total density of 
the universe, as shown by Briel et al. (1992) and White et al. (1993)
in calculating $\frac{M_{gas}}{M_{tot}}=\frac{\Omega_b}{\Omega_{tot}}=
\frac{0.05\pm0.01}{\Omega_{tot}}$.
The determined values of $\Omega_{tot}$ range between 0.2--0.4, and give 
strong evidence for a low density universe. Our values of the gas to
total mass
ratio of 3C 295 suggest $\Omega_{tot}=0.1-0.4$. The errors in the determination
of $\Omega$ come almost exclusively from the uncertainties in $M_{tot}$. 

\subsection{Mass estimates compared to weak lensing results}

Fig.\ref{fig:masse} shows the total  mass profile of 3C 295.\\ 
To be able to compare our results for the mass  with the weak lensing study
by Smail et al. (1997) (-the lensing approach determines the total mass along 
the line-of-sight within a certain radius), we also show the (along the 
line-of-sight) projected mass profile. We project out to a radius of 3~Mpc.\\
There is some discrepancy between the mass results from the lensing analysis
and from the hydrostatic equation used here. The lensing result gives a higher 
mass. The values differ by a factor of less than two. Both estimates have 
relatively large error bars, which, however, do not overlap.\\
Where can this discrepancy come from? Partly it can be due to the fact that
the mass inferred from X-rays is biased downward due to the cooling flow.
The cooling flow in the central parts of the cluster could lower the measured 
overall temperature by ASCA due to cooled gas. Also the small values for
$\beta$ and core radius, typical for a cluster with a central cooling flow
might bias the value for the inferred mass downward
 - the mass calculated from the hydrostatic approach is 
proportional to $\beta$. And finally the fact that we do not see the outer 
regions
of the cluster in X-rays could systematically lower the result of our mass 
determination.
On the other hand the mass determined with the weak
lensing approach is dependent on the mean redshift of the lensed galaxies.
While this is not a strong effect for nearby clusters it plays an important
role  for distant clusters such as 3C 295. 
Smail et al. (1997) used an approach with a mean redshift of the background
galaxies of z=0.83. Increasing the value to for example z=1 lowers the 
inferred mass considerably.  
 An indication that the mass from the
weak lensing might be estimated too high is the fact that the M/L for this 
cluster found by Smail et al.(1997) is relatively high in comparison with 
other clusters in their sample.
Therefore it might be that the different results of the mass can be entirely 
explained by systematic errors in the two approaches. However, we cannot rule
out that there might be some physical reason for this apparent discrepancy
of the mass results, and that for example 3C 295 has either prolate symmetry
along the line-of sight or suffers from projection effect from another 
massive structure.

\section{Conclusions}

In this paper we analyse ROSAT HRI data of the distant cluster 3C 295.
We
fit a one and a two-dimensional isothermal $\beta$-model to the data,
the 
two-dimensional fit taking into account the ellipticity of the cluster and
effects
of the PSF. The one-dimensional model which does not take into account the PSF
gives
higher values for $\beta$ and the core radius $a$; however, the results
agree
within the error bars. To determine the errors of the parameters of the 
two-dimensional fit we use a Monte-Carlo approach.\\
Calculating the cooling time of the central region we confirm previous
findings
by Henry \& Henriksen (1986) of a central cooling flow. We estimate the
cooling
flow radius to lie between 60--120~kpc, depending on the model and adopted
cooling time for the cluster. Generally we find that the isothermal 
$\beta$-model, which does not take into account the effects of the PSF, gives
higher values for the cooling flow radius and the mass accretion
rates.\\
For the determination of the mass accretion rates
we present a simple model, which is solely based on the isothermal 
$\beta$-model, neglecting heat conduction, inhomogenities 
and heat sources.
We show that the inferred mass accretion rates, with values of 
$\dot{M}=400-900$~M$_\odot$/y are dependent on the temperature structure
in
the cooling flow region. A temperature gradient generally lowers the
mass 
accretion rates. \\
The mass accretion rate in the central part of 3C 295 is somewhat
too high for its cooling radius, when compared to the sample of nearby
clusters
of White et al. (1997). As 3C 295 is not the only distant cluster
showing this 
behaviour, this might be an indication that generally this relationship 
between mass accretion rate and cooling flow radius is evolving with
time.
\\
To search for substructure in the cluster we subtract a two-dimensional
cluster
model, following the best fit results of the isothermal $\beta$-model
from the
original data. Apart from a decrement in the South-East in the cluster,
which might be caused by absorption, we do not find strong
indication
for the cluster to deviate from hydrostatic equilibrium.
The fact that 3C 295 seems to be in complete hydrostatic equilibrium,
and has had enough time to form a cooling flow is an 
indication that the cluster must have had considerably time to relax,
which
suggests a low density universe, with or without a cosmological constant.
\\
Calculating the total mass of the cluster based on the hydrostatic
equation
using a temperature estimate of Mushotzky \& Scharf (1997),
we obtain results of M$_{tot}=9.2\pm3.\times 10^{14}$M$_\odot$ at a
radius
of 2.1~Mpc. For the gas to total mass ratio we obtain
M$_{gas}$/M$_{tot}\sim
0.17-0.31$. This ratio is typical for other clusters, and is again a strong 
indication for a low density universe.
\\
\\

I like to thank M. Arnaud, S. Schindler and H. B\"ohringer for very
helpful 
discussions, and Jean-Luc Sauvageot for computer and software support. I like 
to especially thank C.A. Collins for a very careful reading of the manuscript
and the referee, Andy Fabian for his very useful comments.
ROSAT is supported by the BMFT. I thank the CNRS and the MPG for funding.

\hfill
\newpage
\eject 

\begin{figure}
\plotone{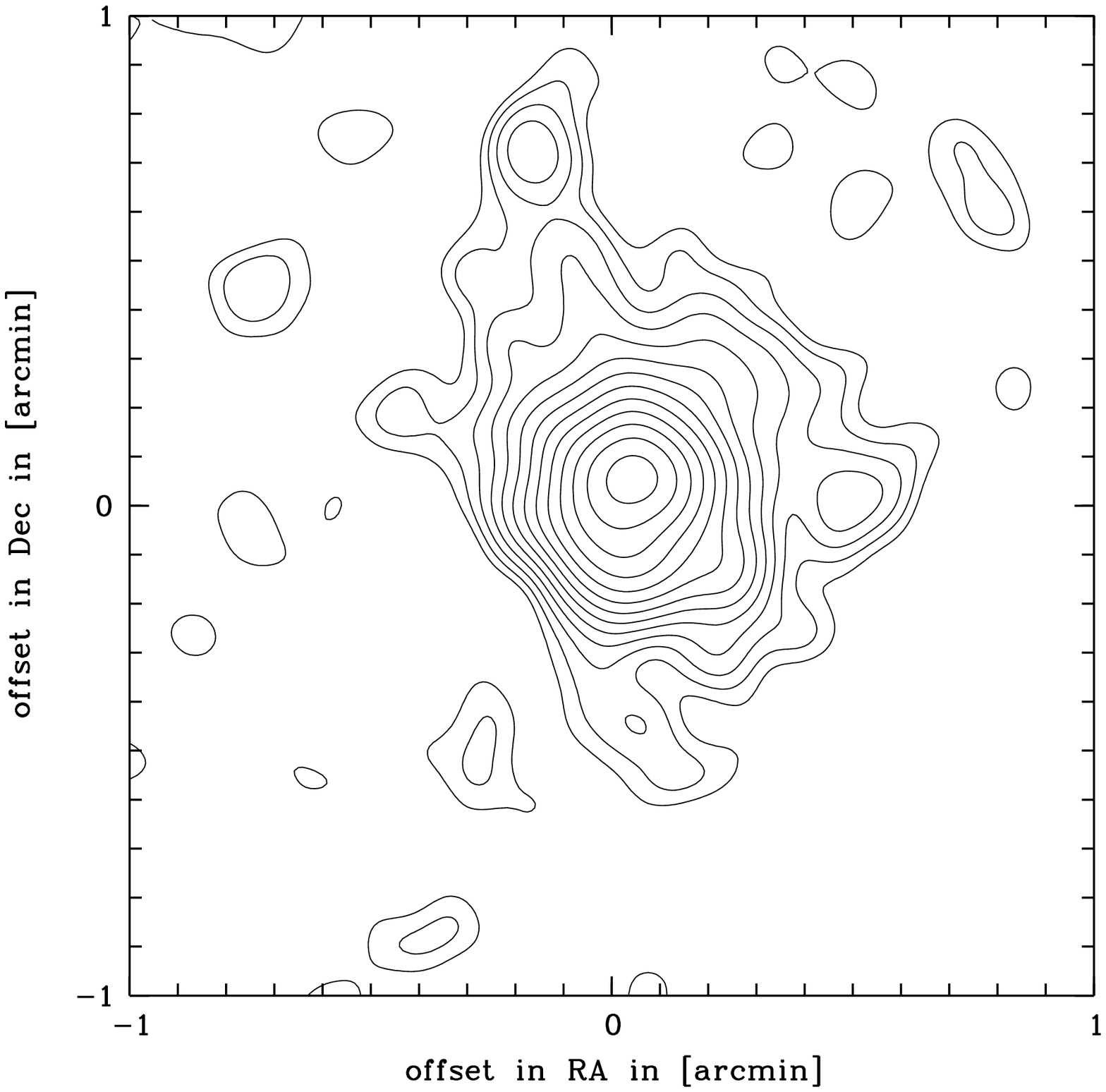}
\caption{}
\label{fig:plot}
\end{figure}

\begin{figure}
\plotone{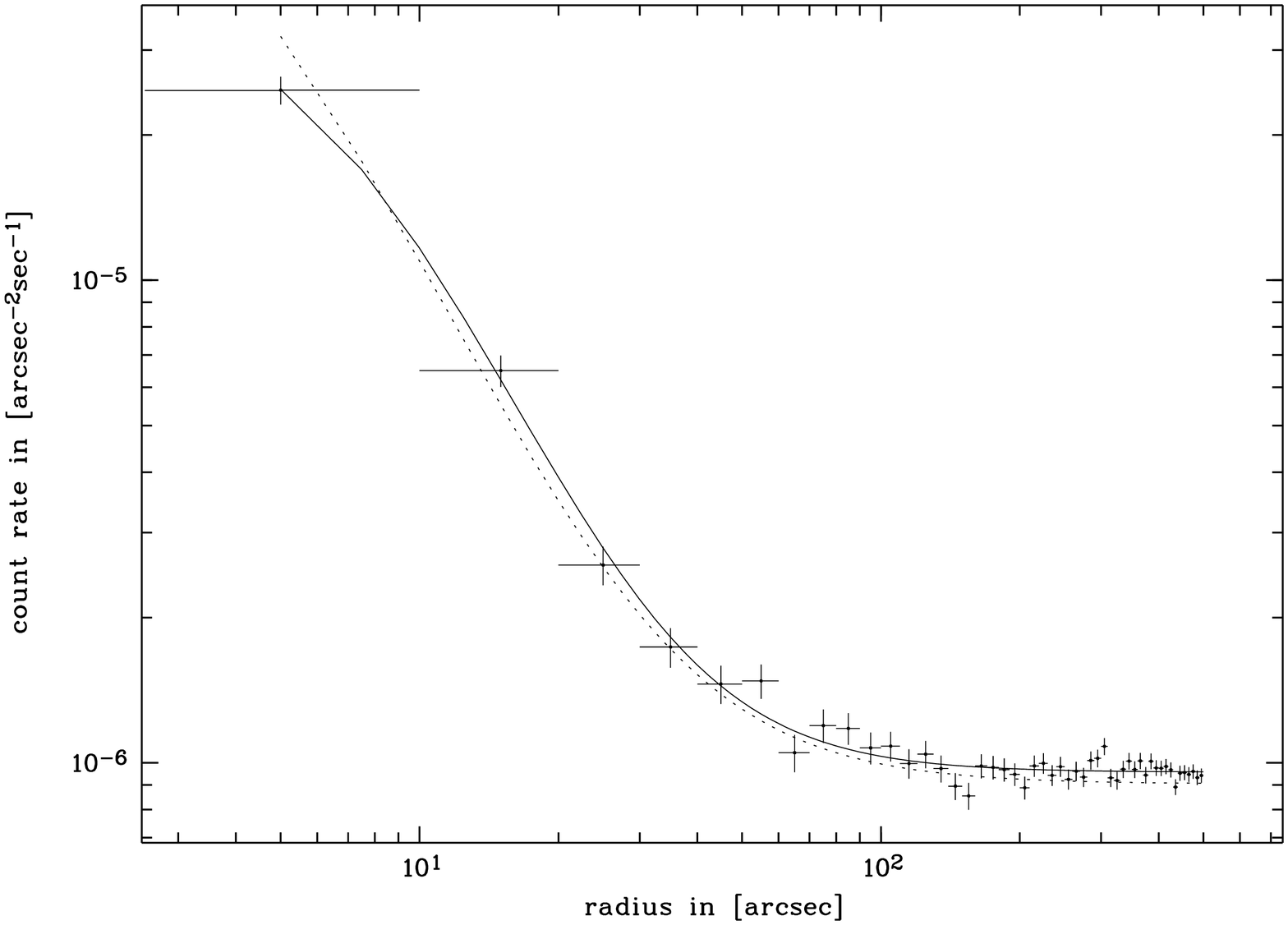}
\caption{}
\label{fig:surf}
\end{figure}

\begin{figure}
\plotone{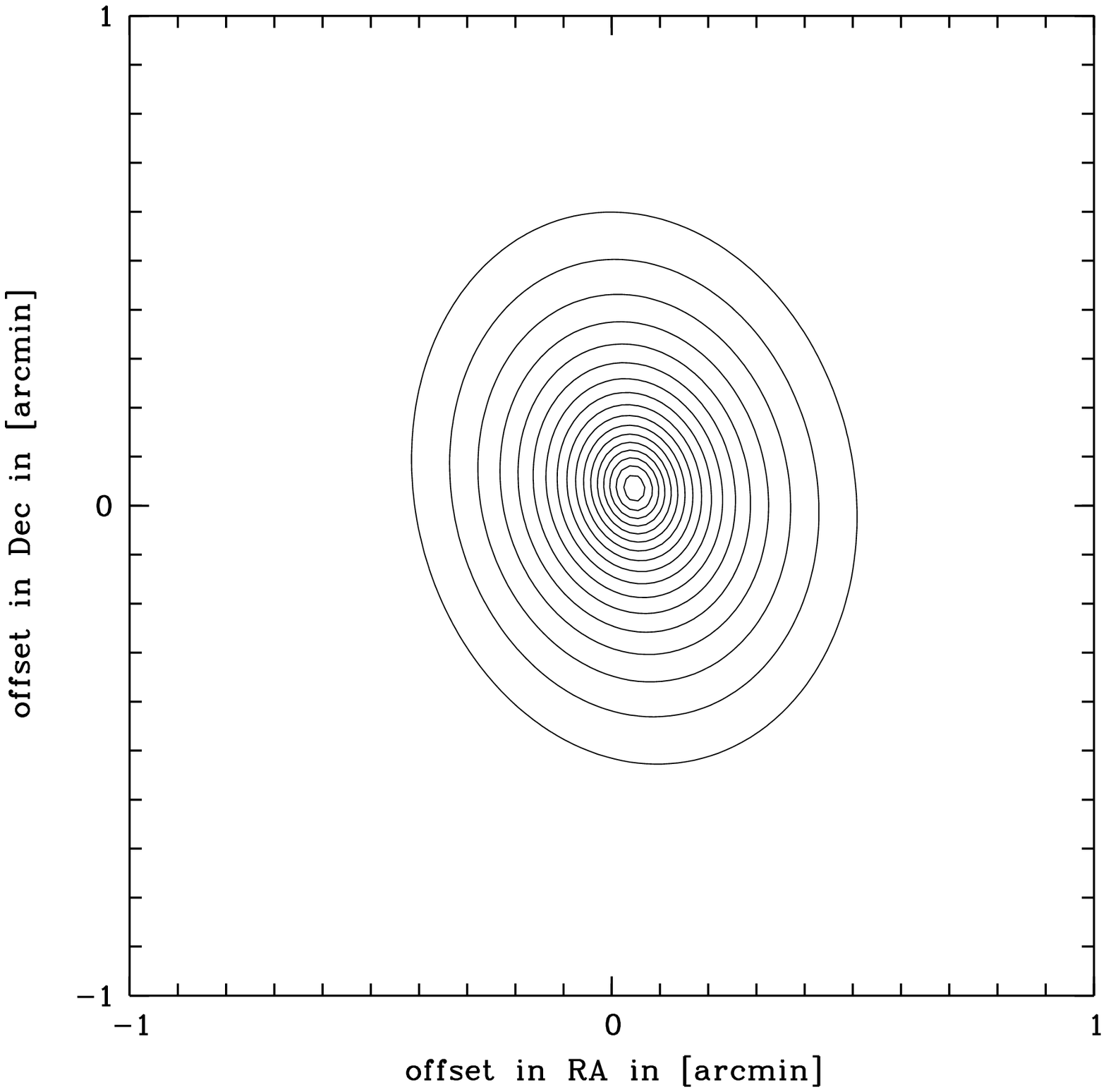}
\caption{}
\label{fig:sym}
\end{figure}

\begin{figure}
\plotone{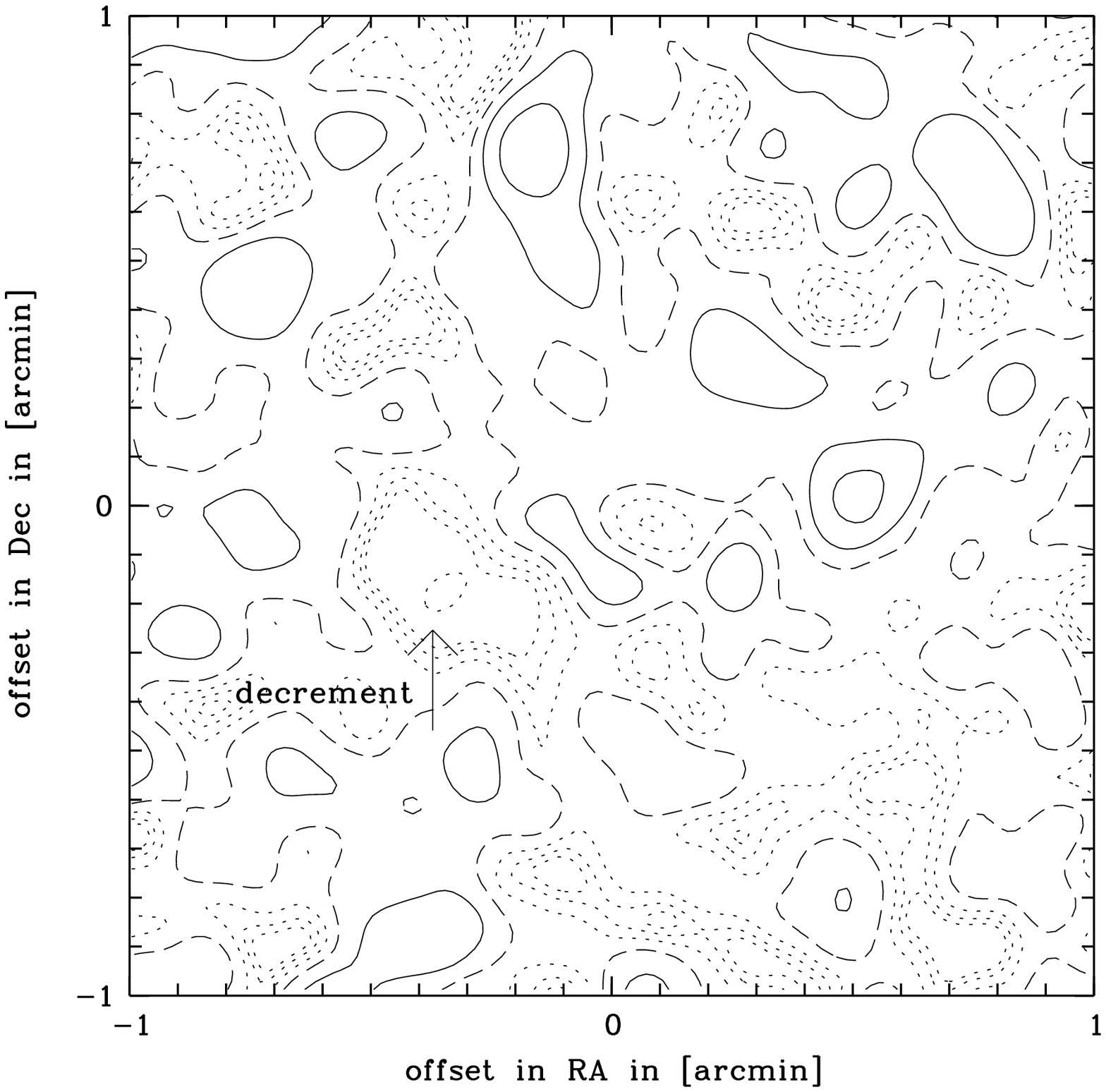}
\caption{}
\label{fig:resi}
\end{figure}

\begin{figure}
\plotone{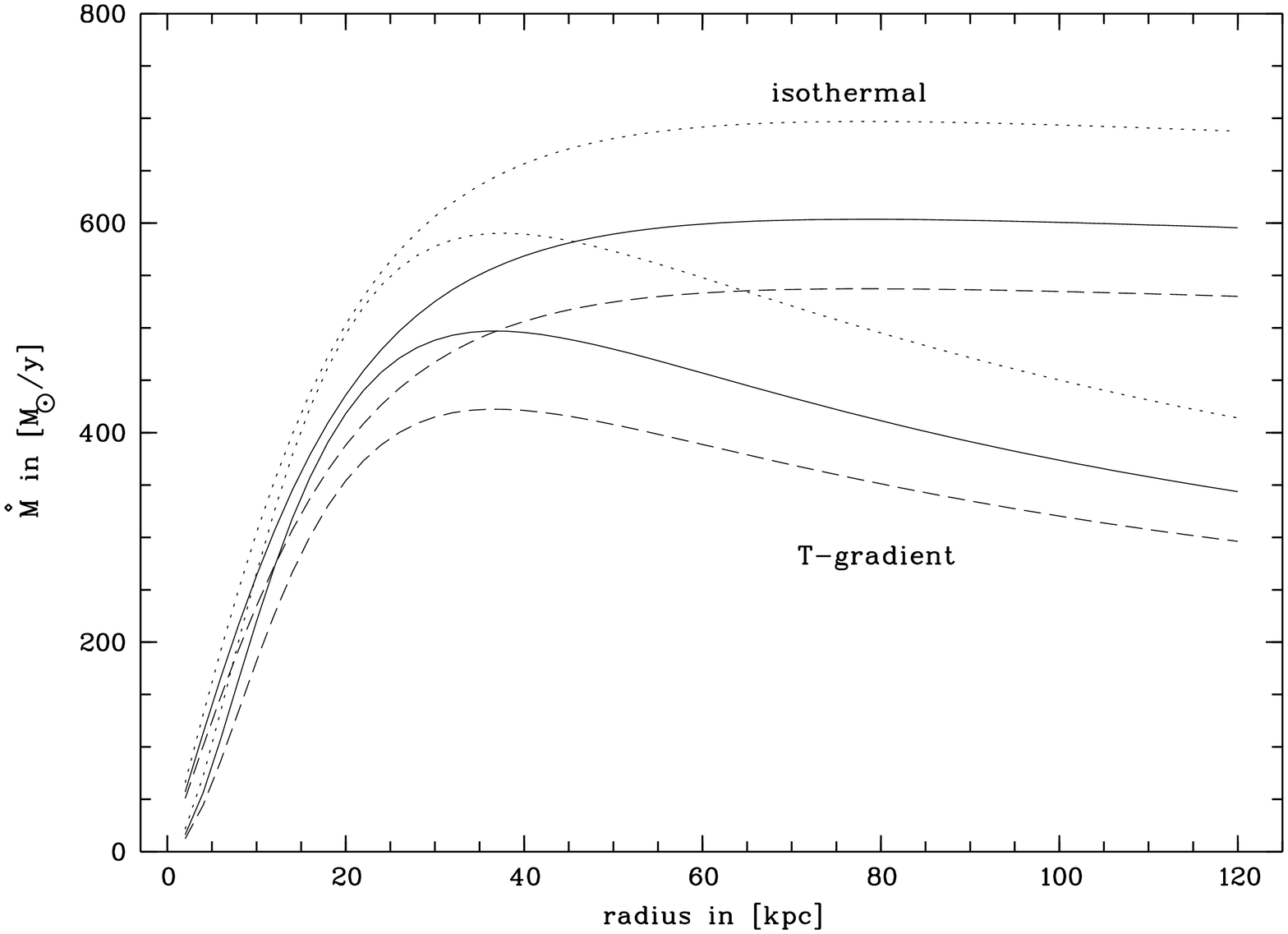}
\caption{}
\label{fig:cf1}
\end{figure}

\begin{figure}
\plotone{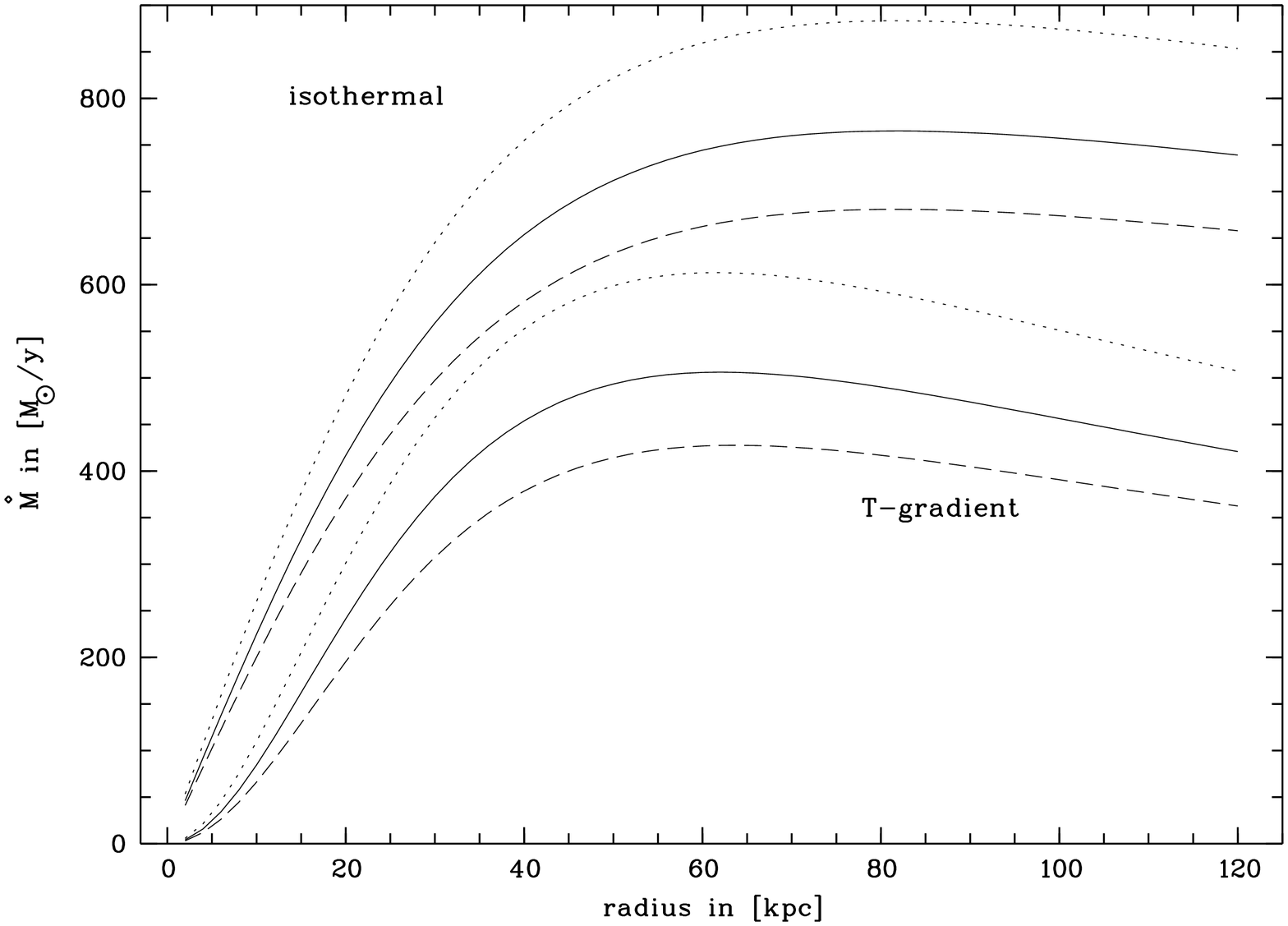}
\caption{}
\label{fig:cf2}
\end{figure}

\begin{figure}
\plotone{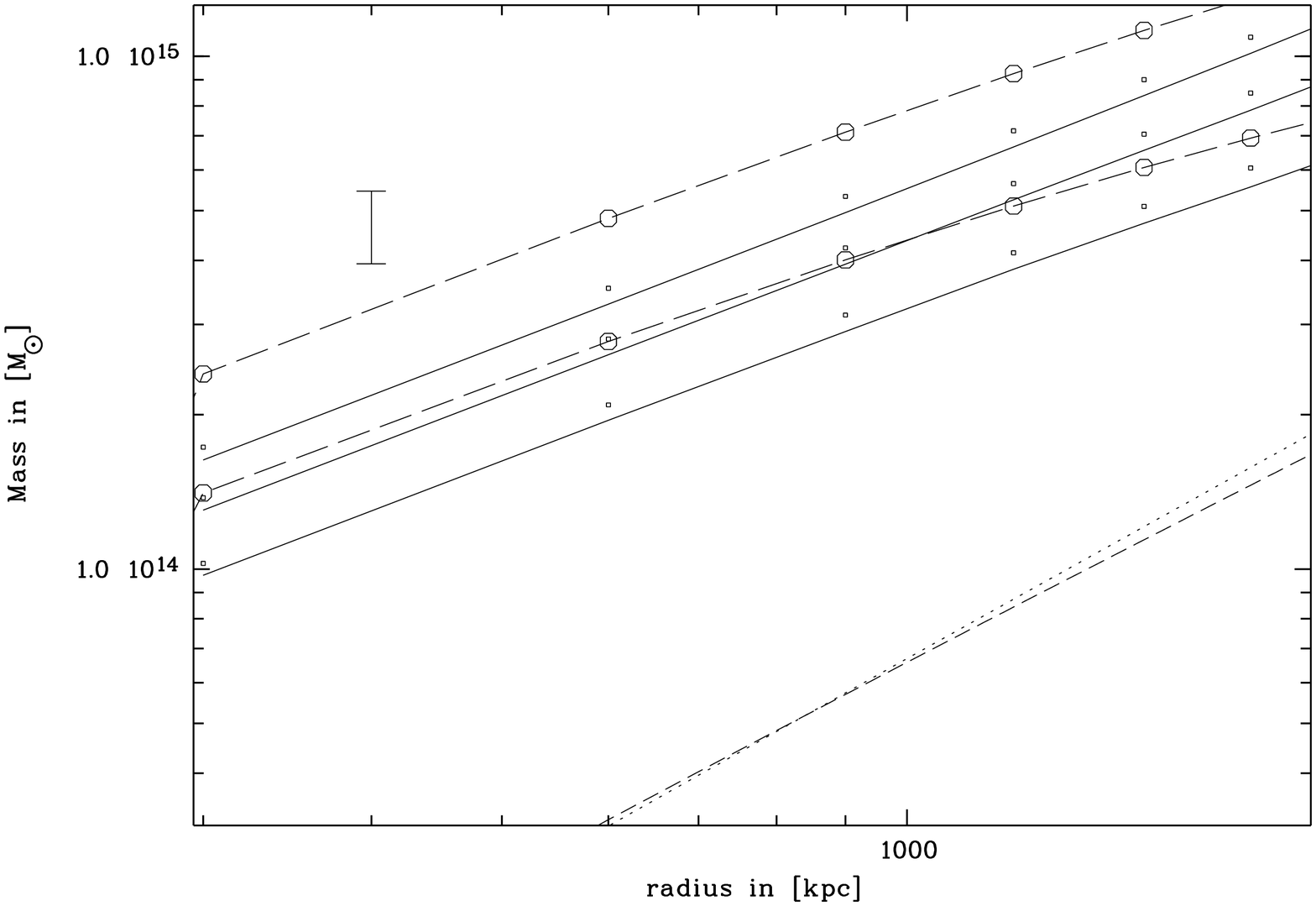}
\caption{}
\label{fig:masse}
\end{figure}

{\bf Fig.}\ref{fig:plot}\\
{
The countrate image of the HRI image of 3C 295. We use only channel 2 to 
9 to optimize the signal to background ratio (for details see text). The
image 
is Gaussian filtered with a $\sigma$ of 3.5 arcsec. The contour levels are
spaced 
logarithmically with $\Delta=0.26$. The lowest contour lies at 7.22 
cts/sec/arcmin$^2$. The image is centered on 
R.A.=$14^{\mbox{h}}11^{\mbox{m}}20.6^{\mbox{s}}$ 
and Dec.=$+52^{\mbox{d}}12^{\mbox{m}}10^{\mbox{s}}$ (J.2000).
}
\\
\\
{\bf Fig.}\ref{fig:surf}\\
{Surface brightness profile of 3C 295. The crosses show the data with 
1-$\sigma$-errors in y-direction. The full line corresponds to the best
fit of 
the 
one-dimensional isothermal $\beta$-fit not taking into account effects
of
the HRI's PSF ($\beta=0.56$). The dotted line shows the results of the
two-dimensional model fit correcting for the PSF (with $\beta=0.52$).
For
displaying we utilize here as core radius the geometrical mean of the
two
core radii, corresponding to 25~kpc$\equiv 3.7$~arcsec.  
}
\\
\\
{\bf Fig.}\ref{fig:sym}\\
{The cluster model corresponding to the two-dimensional fit parameters
shown
in Tab.\ref{tab:beta}. The lowest contour and the spacing are identical
with the
ones in  Fig.\ref{fig:plot}.
}
\\
\\
{\bf Fig.}\ref{fig:resi}\\
{The residuals of the cluster after subtracting the isothermal
$\beta$-model 
following
the best fit results of the two-dimensional fit including background.
The
model is convolved with the PSF before subtraction. The contours show
the significance of the residuals above cluster plus background emission
in
$\sigma$. The dashed line is the zero $\sigma$-line, negative residuals
are
dotted, positive residuals have a full line. The step width is 1
$\sigma$ from
contour to contour.
The shown significance levels take into account the applied
Gaussian filter of 3.5~arcsec.}
\\
\\
{\bf Fig.}\ref{fig:cf1}\\
{
Mass accretion rates in the central part of the cluster using the model
taking
into account the effects of the PSF ($\beta=0.52$, 
$a=\sqrt{a_1a_2}=3.7$~arcsec). Shown are the rates
as a function of radius for the isothermal case (upper lines) 
and for the case of constant
temperature gradient (lower lines). The dotted lines correspond to a 
temperature of 5.8~keV,
 the full line to 7.1~keV, and the dashed line to 9.2~keV. For the
temperature
gradient case we use a central temperature of 2~keV. In this case 5.8,
7.1, and
9.2~keV are the temperatures at a radius of 120~kpc.
}
\\
\\
{\bf Fig.}\ref{fig:cf2}\\
{
The same as previous figure, only difference: we use here the isothermal 
$\beta$-model from the one-dimensional fit, not taking into account the
effects
of the PSF.
}
\\
\\
{\bf Fig.}\ref{fig:masse}\\
{Mass profile of the cluster 3C 295. The full lines correspond to the 
$\beta$-model taking into account PSF effects with $\beta=0.52$ and
a core radius of $a=\sqrt{a_1 a_2}$=3.7~arcsec. The dots
show the result for the model with $\beta=0.56$. Shown are
the mean values $\pm 2\sigma$-errors. The open
circles with the long dashed lines
correspond to the projected mass
profile with $\beta$=0.52 ( - here we only show the error limits 
$\pm 2\sigma$-errors). The dotted line shows the 
gas mass profile with the $\beta=0.52$-model, the dashed line the
profile with
$\beta=0.56$ using a gas temperature of $7.1^{+2.1}_{-1.3}$~keV 
(Mushotzky \& Scharf 1997). The vertical line at 400~kpc shows the
result of 
the weak lensing analysis by Smail et al. (1996) transformed to $H_0=50$. 
} 


\begin{table}[h]
\caption{Serendipitous sources in the FOV of 3C 295}
\begin{tabular}{clccc}
\tableline
\tableline
 & source & RA. & Dec. & total counts \\
\hline
1 & galaxy (z=0.2737)$^1$      & $14^h11^m21.5^s$ & $+52^d12^m53^s$ &
10\\
2 & galaxy (z=0.4733) (Sy1)$^{1 2}$& $14^h11^m23.1^s$ & $+52^d13^m32^s$
& 60\\
3 & QSO (z=1.29)$^{1 3}$           & $14^h11^m19.2^s$ & $+52^d14^m05^s$
& 20\\
\tableline
\multicolumn{5}{l}{The sources agree within  6~arcsec with the X-ray 
position.}\\
\multicolumn{5}{l}{$^1$ Dressler \& Gunn 1992}\\
\multicolumn{5}{l}{$^2$ Hewitt \& Burbidge 1991}\\
\multicolumn{5}{l}{$^3$ Hewitt \& Burbidge 1993}\\
\end{tabular}
\label{tab:exclude}
\end{table}


\begin{table}[t]
\caption{Isothermal $\beta$-fit results}
\begin{tabular}{ccccccccc}
\tableline
\tableline
 & $S_0$ & $a_1$ & $a_2$ & $\beta$ & $x_0$ & $y_0$ & $\alpha$ & $B$ \\
 & [s$^{-1}$arcsec$^{-2}$] & [arcsec] & [arcsec] & & 2000.0 &
2000.0  & 
& [s$^{-1}$arcsec$^{-2}$]\\
\tableline 
1-d & 3.8$\times10^{-5}$ & $7.2^{+3.4}_{-3.3}$ &  &
$0.56^{+0.10}_{-0.07}$ & 
$14^h11^m20.3^s$ & $52^d12^m12^s$ & & $9.6\times 10^{-7}$\\
2-d & $9.2^{+7.0}_{-7.0}\times 10^{-5}$ & $4.2^{+2.3}_{-2.3}$ & 
$3.3^{+1.8}_{-1.8}$ & $0.52^{+0.07}_{-0.07}$ & $14^h11^m20.2^s$  & 
$52^d12^m12^s$ & $14^{+27 \circ}_{-27}  $ & $9.1^{+0.09}_{-0.10}\times 
10^{-7}$ \\
\tableline 
\end{tabular}
\label{tab:beta}
\end{table}


\begin{table}[h]
\caption{Fit results of the simulated cluster model}
\begin{tabular}{lcccccccc}
\tableline
\tableline
 & $S_0$ & $a_1$ & $a_2$ & $\beta$ & $x_0$ & $y_0$ & $\alpha$ & $B$\\
 & [s$^{-1}$arcsec$^{-2}$] & [arcsec] & [arcsec] & & [arcsec] &
[arcsec]  & 
& [s$^{-1}$arcsec$^{-2}$]\\
\tableline
input & $9.2\times10^{-5}$ & 4.2 & 3.3 & 0.52 & 0.0 & 0.0 & $14^\circ$ & 
$9.0\times10^{-7}$ \\
fit & $10.5^{5.8}_{-5.8}\times10^{-5}$ & $4.0^{+2.0}_{-2.0}$ & 
$3.2^{+1.5}_{-1.5}$ & $0.52^{+0.05}_{-0.05}$ & $0.0^{+0.7}_{-0.7}$ 
& $0.0^{+0.8}_{-0.8}$ & $14^{+28\circ}_{-28}$ & 
$9.1^{+0.7}_{-0.7}\times10^{-7}$  \\
\end{tabular}
\label{tab:simul}
\end{table}


\begin{table}[h]
\caption{Cooling rates - adopted from B\"ohringer \& Hensler (1989).}
\begin{tabular}{ll}
\tableline
\tableline
$\Lambda(T)$ & $k_b$T \\

 [10$^{-23}$erg s$^{-1}$ cm$^6$] &  [keV]\\
\tableline
1.4 & 2.0 \\
2.1 & 5.8 \\
2.3 & 7.8 \\
2.6 & 9.2 \\ 
\tableline
\end{tabular}
\label{tab:lambda}
\end{table}


\begin{table}[h]
\caption{Results of the Monte Carlo mass analysis}
\begin{tabular}{lllll}
\tableline
\tableline
 & total mass ($\beta=0.52$) & total mass ($\beta=0.56$) & gas mass 
($\beta=0.52$) & gas mass ($\beta=0.56$)\\
 & [$10^{14}$M$_\odot$] & [$10^{14}$M$_\odot$] & [$10^{14}$M$_\odot$] &
[$10^{14}$M$_\odot$] \\
\tableline
at 2.1~Mpc & $9.2\pm2.7$ & $9.9\pm3.0$ & 2.0 & 1.8\\  
\tableline
\end{tabular}
\label{tab:masse}
\end{table}

\hfill
\newpage
\eject

\begin{thebibliography}{}

\bibitem[] {} Akujor C.E., L\"udke E., Browne I.W.A., Leahy J.P.,
Garrington
S.T., Jackson N.,\& Thomasson P 1994, A\&AS, 105, 247
\bibitem[] {} Allen S.W., Fabian A.C., Edge A.C., Bautz M.W.,\& Furuzawa
A.
1996a, MNRAS, 283, 263
\bibitem[] {} Allen S.W., Fabian A.C.,\& Kneib J.P. 1996b, MNRAS, 279,
615
\bibitem[] {} B\"ohringer H.,\& Hensler G. 1989, A\&A, 215, 147
\bibitem[] {} B\"ohringer H., Voges W., Fabian A.C., Edge A.C.,\&
Neumann D.M.
1993, MNRAS, 264L, 25
\bibitem[] {} B\"ohringer H. 1994, in Proceedings of the NATO Advanced
Study
Institute on Cosmological Aspects of X-ray Clusters of Galaxies, ed. W. 
Seitter (Boston: Kluwer), 123
\bibitem[] {} B\"ohringer H., Tanaka Y., Mushotzky R.F., Ikebe Y.,\&
Hattori M.
1998, A\&A, in press 
\bibitem[] {} Briel U.G., Henry J.P.,\& B\"ohringer H. 1992, A\&A, 259,
L13
\bibitem[] {} Cavaliere A.,\& Fusco-Femiano R. 1976, A\&A 49, 137
\bibitem[] {} Cavaliere A.,\& Fusco-Femiano R. 1981, A\&A 100, 194
\bibitem[] {} Cen R.,\& Ostriker J.P. 1993, ApJ, 417, 404
\bibitem[] {} Cowie L.L.,\& Binney J. 1977, ApJ, 215, 723 
\bibitem[] {} David L.P. et al. 1997, The ROSAT High Resolution Imager
(HRI)
Calibration Report, U.S. ROSAT SCIENCE DATA CENTER/SAO
\bibitem[] {} Dickey J.M.,\& Lockman F.J. 1990, A\&AR, 28, 215
\bibitem[] {} Dressler A.,\& Gunn J.E., 1992, ApJS, 78, 1
\bibitem[] {} Edge A.C., Fabian A.C., Allen S.W., Crawford C.S., White
D.A.,
B\"ohringer H.,\& Voges W. 1994, 270, L1
\bibitem[] {} Evrard A.E., Metzler C.A.,\& Navarro J.N., 1996, ApJ, 469,
494
\bibitem[] {} Evrard A.E. 1990, 363, 349
\bibitem[] {} Fabian A.C.,\& Nulsen P.E.J. 1977, MNRAS, 180, 479 
\bibitem[] {} Fabian A.C., Nulsen P.E.J.,\& Canizares C.R. 1984, Nat.
310, 733
\bibitem[] {} Fabian A.C., Nulsen P.E.J.,\& Canizares C.R. 1991, A\&AR,
2 191
\bibitem[] {} Fabian A.C.,\& C.S. Crawford 1995, MNRAS, 274, L63
\bibitem[] {} Harris D.E., Silverman J.D., Hasinger G., Lehmann I. 1998,
A\&AS, 133, 431
\bibitem[] {} Henry J.P.,\& Henriksen M.J. 1986, ApJ, 301, 689
\bibitem[] {} Hewitt A.,\& Burbidge G. 1991, ApJS, 75, 297
\bibitem[] {} Hewitt A.,\& Burbidge G. 1993, ApJS, 87, 451 
\bibitem[] {} Jones C.,\& Forman W. 1992, Clusters and Sperclusters of 
Galaxies,
ed. A.C. Fabian (Dordrecht: Kluwer), 49
\bibitem[] {} Lea S.M., Silk J., Kellog E.,\& Murray S. 1973, ApJ, 184,
L105
\bibitem[] {} Lea S.M.,\& Henry J.P. 1988, ApJ, 332, 81
\bibitem[] {} Mathews W.G.,\& Bregman J.N. 1978, ApJ, ApJ, 244, 308
\bibitem[] {} Morse J.A. 1994, PASP, 106, 675
\bibitem[] {} Mushotzky R.F.,\& Scharf C.A. 1997, ApJ, 482, L13 
\bibitem[] {} Neumann D.M.,\& B\"ohringer H. 1995, A\&A, 301, 865 
\bibitem[] {} Neumann D.M.,\& B\"ohringer H. 1997, MNRAS, 289, 123
\bibitem[] {} Neumann D.M.,\& B\"ohringer H. 1998, ApJ, submitted 
\bibitem[] {} Pierre M., Le Borgne J.F., Soucail G.,\& Kneib J.P. 1996
A\&A, 
311 413
\bibitem[] {} Sarazin C. 1986, Rev. Mod. Phys. 58, 1
\bibitem[] {} Sarazin C. 1988, X-ray emission from Clusters of Galaxies,
C.U.P.
\bibitem[] {} Schindler S., 1996, A\&A, 305, 756
\bibitem[] {} Schindler S., Hattori M., Neumann D.M.,\& B\"ohringer H.
1997, 
317, 646
\bibitem[] {} Schindler S.,\& Prieto M.A. 1997, A\&A, 327, 37
\bibitem[] {} Smail I., Ellis R.S., Dressler A., Couch W.J., Oemler
A.Jr.,
Sharples R.M.,\& Butcher H. 1997, ApJ, 479, 70
\bibitem[] {} Squires G., Kaiser N., Babul A., Fahlman G., Woods D.,
Neumann D.M.,\& B\"ohringer H. 1996, ApJ, 461 572
\bibitem[] {} Squires G., Neumann D.M., Kaiser N., Arnaud M., Babul A.,
B\"ohringer H., Fahlman G.,\& Woods G. 1997, ApJ, 482, 648
\bibitem[] {} Thimm G.J., R\"oser H.-J., Hippelein H.,\& Meisenheimer K.
1994,
A\&A, 285, 785
\bibitem[] {} Thomas P.A., Fabian A.C.,\& Nulsen P.E.J. 1987, MNRAS,
228, 973 
\bibitem[] {} Walker T.P., Steigman G., Kang H., Schramm D.M.,\& 
Olive K.A. 1991, ApJ, 376, 51
\bibitem[] {} White D.A., Jones C.,\& Forman W. 1997, MNRAS, 292, 419
\bibitem[] {} White S.D.M., Navarro J.F., Evrard A.E.,\& Frenk C.S.
1993, Nat. 366, 429 

\end{thebibliography}
\end{document}